\documentclass[conference]{IEEEtran}
\IEEEoverridecommandlockouts
\usepackage{cite}
\usepackage{amsmath,amssymb,amsfonts}
\usepackage{algorithmic}
\usepackage{graphicx}
\usepackage{textcomp}
\usepackage{subcaption}
\usepackage{xcolor}
\def\BibTeX{{\rm B\kern-.05em{\sc i\kern-.025em b}\kern-.08em
    T\kern-.1667em\lower.7ex\hbox{E}\kern-.125emX}}
\begin{document}

\title{Synthetic Dynamic PMU Data Generation: A Generative Adversarial Network Approach\\
{\footnotesize }
}

\author{\IEEEauthorblockN{Xiangtian Zheng, Bin Wang, Le Xie}
\IEEEauthorblockA{\textit{Dept. of Electrical Engineering} \\
\textit{Texas A{\&}M University}\\
College Station, USA \\
$\lbrace$zxt0515, binwang, le.xie$\rbrace$@tamu.edu}
}
\maketitle

\begin{abstract}
This paper concerns with the production of synthetic phasor measurement unit (PMU) data for research and education purposes. Due to the confidentiality of real PMU data and no public access to the real power systems infrastructure information, the lack of credible realistic data becomes a growing concern. Instead of constructing synthetic power grids and then producing synthetic PMU measurement data by time simulations, we propose a model-free approach to directly generate synthetic PMU data. we train the generative adversarial network (GAN) with real PMU data, which can be used to generate synthetic PMU data capturing the system dynamic behaviors. To validate the sequential generation by GAN to mimic PMU data, we theoretically analyze GAN's capacity of learning system dynamics. Further by evaluating the synthetic PMU data by a proposed quantitative method, we verify GAN's potential to synthesize realistic samples and meanwhile realize that GAN model in this paper still has room to improve. Moreover it is the first time that such generative model is applied to synthesize PMU data.
\end{abstract}


\section{Introduction}
This paper is motivated by the need to have credible synthetic dynamic data for research and education in large power systems. Given the fact that most of the real-world measurement data are confidential and subject to Critical Energy/Electric Infrastructure Information (CEII) protected by the Energy Act, there is a growing concern to have credible realistic synthetic data for training and research purposes.

The demand of data is usually satisfied by the traditional simulation based methods in many power system applications and designs. However, due to the security reasons, the system topology and the parameters of the facilities are usually not freely shared among involved participants. Data simulated from the widely used models such as IEEE standard models may show the system behavior different from that of a real power system. Proposed as a solution to this concern, synthetic grid\cite{b13} provides a natural simulation platform via a large number of power grid test cases with realistic topologies, scalable network size and realistic electrical parameter settings. Although simulation data via this way represent many properties of real power systems, this method may require time-consuming extensive offline simulation of many contingencies that may or may not fully capture the dynamics of the real world. Building physical models usually faces the problem of model accuracy and the limitation caused by the model formula. Considering this, a model-free method would be a promising alternative to generate additional synthetic PMU data to mimic the behavior of real power system by utilizing the available real PMU data.

To synthesize realistic data directly, deep generative models. Generative Adversarial Nets (GAN)\cite{b1} became one of the most popular unsupervised methods since proposed in 2014. The main idea of GAN is to train generative models via an adversarial process where two models are trained simultaneously: a generative model G that captures the mapping from a known distribution to the data distribution and a discriminative model D that estimates the probability that a sample comes from training data set rather than generated by G. These two players aim to minimize their own costs and the final state of this adversarial game is the Nash equilibrium. In the original version, GAN faced problems like vanishing gradients, mode collapse and training instability. Therefore, substantial theoretical research works were conducted to address them, among which Wasserstein GAN\cite{b2}\cite{b3}, Improved Wasserstein GAN\cite{b4} and Least Square GAN\cite{b5} are the famous representatives.

Although there still exists room to improve the theory of GAN, researchers cannot be stopped from diving deeply and enthusiastically into possible applications and GAN has already achieved a great success especially in the field of computer vision. A modified version of GAN, i.e. CycleGAN\cite{b6}, transferred images from one domain to another domain, for instance, from real sceneries to Van Gogh style paintings. The goal of such image-to-image translation is to learn the mapping from input images to output images by using a training set of paired images. Other researchers showed that it is possible to use descriptive language to generate corresponding images by a proposed method of text translation to images\cite{b7}.  Another complementary learning strategy\cite{b8}, mainly based on GAN, was proposed to build the internal representation of videos and hence learn to predict the following pictures for the given video sequence. It was illustrated that such method could model the content and dynamics of images in the video precisely. Not limited in the image and video generation, GAN extends its magic power to other industry as well. AnoGAN\cite{b9}, a deep convolutional generative adversarial network, was trained to capture the disease progress. By mapping from the image space to a latent space, this model could play the role of a marker and provide anomaly scoring for each tomography image of the retina.  MIDINET\cite{b10} was proposed for symbolic-domain music generation by a novel conditional mechanism. The final trained model could be expanded to generate multiple MIDI channels. Particularly for the applications of power systems, the only representative is the work of scenario generation\cite{b11}. It was demonstrated that the proposed method could generate realistic looking wind and photovoltaic power profiles with a rich diversity. That paper also illustrated the way how to generate scenarios based on the different conditions of interest as well. However, by now, it remains blank to realize the idea of applying GAN to synthesize PMU data.

The main contributions of this paper are as follows: 1) Through the theoretical analysis, we show the potential of  GAN to learn the system dynamic behaviors from given PMU measurement data and generate additional synthetic PMU data; 2) We propose a quantitative method to evaluate how realistic these synthetic samples are.

The paper is organized as follows: Section II mainly introduces the basics of the adversarial game in GAN and its modified versions. Section III presents the generation and validation of synthetic PMU data using Wasserstein GAN to learn the underlying system dynamic behavior from given PMU data. Section IV draws conclusions and states the future work.   

\section{Review of GAN}
In this section, we introduce the GAN applied in this paper by reviewing the basic idea, objective function as well as detailed training procedure. 

\subsection{Basic idea of GAN}
GAN was firstly proposed in 2014 and its well-designed adversarial game has shown the great performance in various field, especially for image generation. The architecture of GAN is shown in the figure 1. The basic idea behind GAN is that if the distribution of the real and synthetic samples are the same, then synthetic data are supposed to be adequately realistic. Denote the true distribution of real data by $P_r$. Suppose we have noise data set $Z$ subject to a known distribution such as jointly Gaussian distribution, denoted by $P_z$. The goal is to find a function mapping from $P_z$ to $P_r$. It is accomplished by training two models alternately: the generative model and the discriminative model.

\begin{figure}[htbp]
\centerline{\includegraphics[scale=0.6]{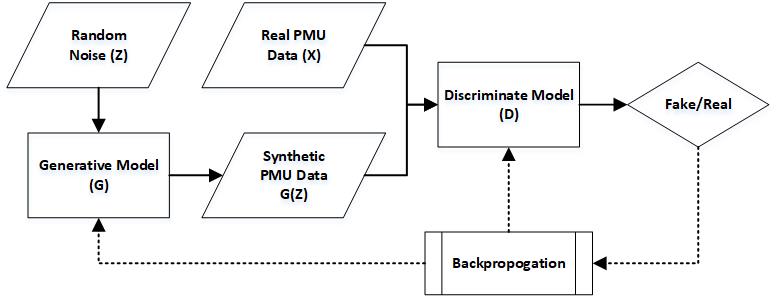}}
\caption{Diagram of GAN}
\label{fig}
\end{figure}

Generative model $G$ is trained to be the function that maps the distribution as mentioned above. Usually it is achieved by the deep neural network because of its capacity of mimicking complex functions. The input data $Z$ are easily sampled from the known distribution $P_{Z}$. Denote the model parameters as $\theta_{G}$ and the model as $G(Z|\theta_{G})$. The distribution of the synthetic data $X_{G}$ generated by $G(Z|\theta_{G})$ is denoted as $P_{g}$. Given a batch of noise data as input, the generative model is trained to generate realistic data and the synthetic distribution $P_g$ is supposed to as close to $P_r$ as possible.

Discriminative model $D$ is achieved by the deep neural network as well and parameterized by $\theta_{D}$. Taking samples from either of real data set $X_r$ and synthetic data set $X_G$, its output values indicate how close these two distributions are. The job of the discriminative model $D$ is to distinguish $P_r$ from $P_g$ by maximizing the difference between the outputs for real and synthetic samples.

\subsection{Objective Function}
With the descriptive definition of these two models, the next step is to formulate their objective functions that determine the rule of this adversarial game. Firstly it is necessary to introduce a criterion that evaluates the distance between two distributions, called Wasserstein distance. Its mathematical expression is shown below, where $\Pi(P_r,P_g)$ denotes the set of all joint distributions whose marginal distributions are respectively $P_r$ and $P_g$. The lower bound 0 is reached only when the distributions are the same.
\begin{equation}
W(P_r,P_g)=inf_{\gamma\in\Pi(P_r,P_g)}(E_{(x,y)\sim\gamma}[||x-y||])
\end{equation} 

However, the objective function in such formula is impractical to be achieved by the neural networks. Instead, ts duality offers the embryonic form of the objective function, where the function $f$ is a 1-Lipschitz function. To obtain the Wasserstein distance, the supremum is searched over the space of all the 1-Lipschitz functions.  
\begin{equation}
W(P_r,P_g)=sup_{|f|_L\leq{1}}(E_{x\sim{P_r}}[f(x)]-E_{x\sim{P_{g}}}[f(x)])
\end{equation}

If transforming the function above by replacing $f$ by $D$ model and  $x$ subject to $P_{g}$ by $G(z)$, then we obtain the objective function of $D$ model and some technique will be introduced later to guarantee $D$ model is always a 1-Lipschitz function during the training. 
\begin{equation}
\underset{D}{Max}\; V(D)=\mathbb{E}_{x\sim{P_r}}[D(x)]-\mathbb{E}_{z\sim{P_{z}}}[D(G(z))]\label{eq}
\end{equation}
The value of this objective function means the Wasserstein distance between the distributions of the real data and synthetic samples if given a $G$ model.

Since the task of $G$ model is to synthesize the realistic data, its training target is to minimize the Wasserstein distance. Therefore the total objective function comes naturally.
\begin{equation}
\underset{G}{Min} \underset{D}{Max}\; V(G,D)=\mathbb{E}_{x\sim{P_r}}[D(x)]-\mathbb{E}_{z\sim{P_{z}}}[D(G(z))]\label{eq}
\end{equation}  

\subsection{Training Procedure}
With the defined objective function, the detailed training procedure of this minimax game follows: 

1) randomly sampling n samples from real data set and n samples from noise data set; 

2) repeating k times of updating D model with the backpropagation according to the following equation:
\begin{equation}
\nabla_{\theta_D}\frac{1}{n}\sum_{i=1}^{n}[-D(x^i)+D(G(z^i))]
\label{eq}
\end{equation}
After each iteration, the parameters in D model will be clipped within a predefined range $[-c,c]$. Clipping is to constrain D model to always be a 1-Lipschitz function during the training. 

3) repeating 1 time of updating G model with the backpropagation according to the following equation:
\begin{equation}
\nabla_{\theta_G}\frac{1}{n}\sum_{i=1}^{n}[-D(G(z^i))]
\label{eq}
\end{equation}

4) iteratively executing the steps above until the iteration step reaches the preset number. 

In this paper, Wasserstein distance will be used in sequential generation of PMU data. Since the minimax objective function in Wasserstein GAN can be interpreted as the dual of the Wasserstein distance, it is a continuous function of evaluating the distance between two distributions and hence it guarantees the gradients always exist. Such characteristic improves model's robustness. Once the minimax game converges, the objective function value is suppose to be 0, meaning the synthetic distribution is the same as the real one. 

\section{Synthetic PMU Data Generation Using GAN}
\subsection{Problem Clarification}
For the case of power systems, if ignoring radom noise, internal relationship of a time series measured by PMU usually follows the underlying physical model represented by a set of differential-algebraic equations. Our objective is to train GAN to synthesize realistic time series that follows a physical model in the specific formula. In other words, $G$ model is expected to learn the system dynamic behaviors from the training data set. Note that we never introduce any priori knowledge of power system models during the training.

To verify that GAN has the capability of learning the dynamic behaviors, it is necessary to clarify what the dynamic behaviors are. For a given power system, its dynamic behavior $X$ can be represented by a function $F$ such that
\begin{equation}
X=F(x_0|\theta)\label{eq}
\end{equation}
where $\theta$ represents the system parameters, $x_0$ is an $n$-dimension state vector representing the initial condition, and $X$ is an $n\times m$ matrix that contains $n$-channel of $m$-step sequences and starts from the initial state $x_0$.  If $\theta$ is fixed, then $F$ is a deterministic function. Suppose $x_0$ is uniformly distributed in a fixed and bounded range, then the range where $X$ distributes in the high dimensional space is determined by $F$. The probability of one sample $X$ depends on that of the corresponding $x_0$. Therefore, we can say the distribution $P_X$ of $X$ is determined by $F$. Obviously only two same systems have the same $P_X$. Further the concept of the dynamic behavior can be mathematically described by the distribution $P_X$. Hence comparing the distributions is equivalent to evaluating the similarity of the dynamic behaviors. Since GAN was proposed to mimic the distribution of the training data set, it surely has the capability of learning the system dynamic behaviors if given enough data. In this paper our objective is to let the distribution of $G$ model's output as close to that of training data as possible. Note that since the input of GAN is random noise rather than initial states, what $G$ model mimics is not the dynamic function $F(\cdot|\theta)$, which will be left for the future work.
 
\subsection{Data Preparation and Training Details}
In the experiment case of sequential generation, we aim to train Wasserstein GAN to learn the underlying dynamic behavior of time series samples and use it to synthesize realistic sequences. The simulation models used for producing training samples are respectively a single-machine-infinite-bus (SMIB) power system and the IEEE 9-bus system as shown in the figure 4 and 5. In the case of SMIB system model, initial rotor speed and phase of the generator are selected from reasonable ranges, while other parameters of the generator and transmission line are fixed. Various initial states can be regarded as the state after disturbances and following profiles reflect the post-disturbance system dynamics. In the case of IEEE 9-bus system, the location of the fault is randomly set on either of buses and its duration is randomly selected from a reasonable range as well. Time series recording starts from the moment when the fault disappears and lasts for 4 seconds with the sampling frequency of 60 Hz. Since this section aims to show Wasserstein GAN's capability of learning the internal temporal features, only PMU equipped on the transmission line between bus 4 and 6 is recorded to collect current measurements as training data samples.
\begin{figure}[htb]
\centering
\includegraphics[scale=0.6]{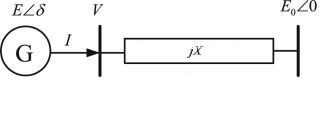}
\caption{Single-machine-infinite-bus power system}\label{fig}
\end{figure}
\begin{figure}[htb]
\centering
\includegraphics[scale=0.3]{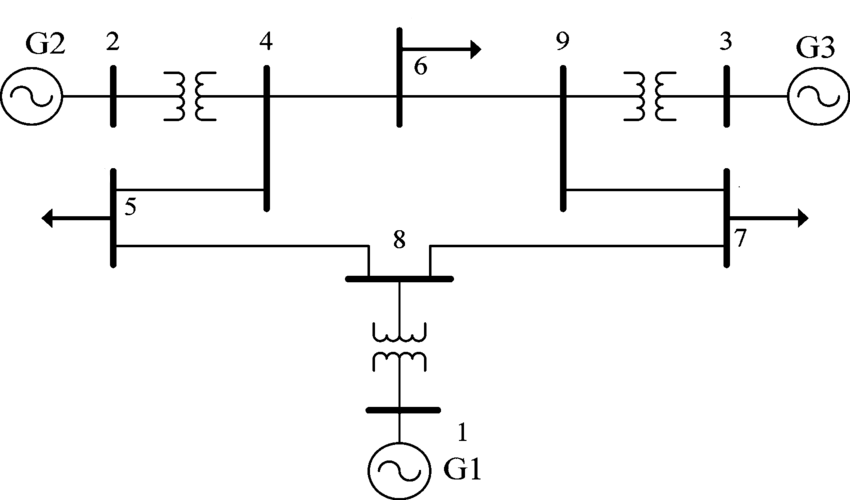}
\caption{IEEE 9 bus power system}\label{fig}
\end{figure}

The input data of $G$ model are random noise samples that are subject to the Gaussian distribution. The training data samples, as part of the input of $D$ model, contain tuples of multiple time series, collected from the first 200 time steps of the data prepared above. In the following experiment, GAN focuses on synthesizing the profiles of current measurements including current magnitude and phase. Therefore, the input of $D$ model have double tuples.

In this paper, both of $G$ and $D$ models are achieved by the standard neural networks, as introduced in Section II, which turn out to work well. GAN models achieved by convolutional neural networks or recurrent neural networks may help improve the performance and they are left to be investigated in our future work. $G$ model internally contains 2 parallel fully connected neural networks to synthesize the magnitude and phase profiles respectively. $D$ model has 2 fully connected layers. The output layers use sigmoid functions, while all other layers use ReLU activation functions. During the training process, each iteration contains 5 times of updating $D$ model and 1 time of updating $G$ model. The learning rate of both models is $10^{-4}$. 

Usually the output of neural networks have high frequency components that cannot be entirely eliminated by modifying the neural networks or adding the extra penalty. Therefore, the final synthetic samples are obtained by filtering the raw synthetic samples with a low-pass filter. 
\subsection{Synthetic Results and Validation}
With the training details introduced in the last subsection, we trained Wasserstein GAN to generate a synthetic time series. Figure 6 and 7 (Please make plots a and c, b and d respectively have identical scales in y-axis) respectively illustrate typical real samples and realistic synthetic samples in the SMIB system and IEEE 9-bus system.

\begin{figure}[t]
\centering
\begin{subfigure}[t]{0.45\linewidth}
    \includegraphics[scale=0.2]{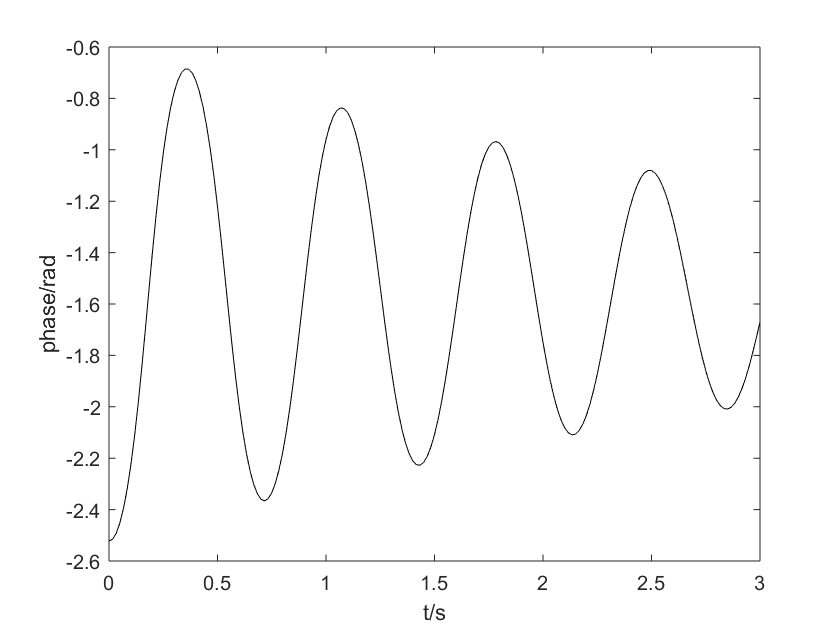}
    \caption{real current phase}
\end{subfigure}
\begin{subfigure}[t]{.45\linewidth}
    \includegraphics[scale=0.2]{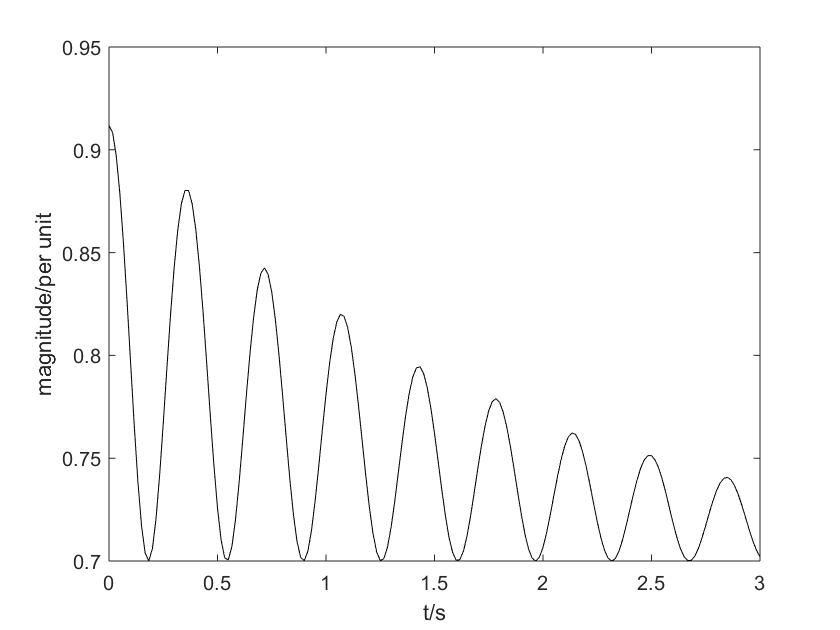}
    \caption{real current magnitude}
\end{subfigure}\\
\begin{subfigure}[t]{0.45\linewidth}
    \includegraphics[scale=0.2]{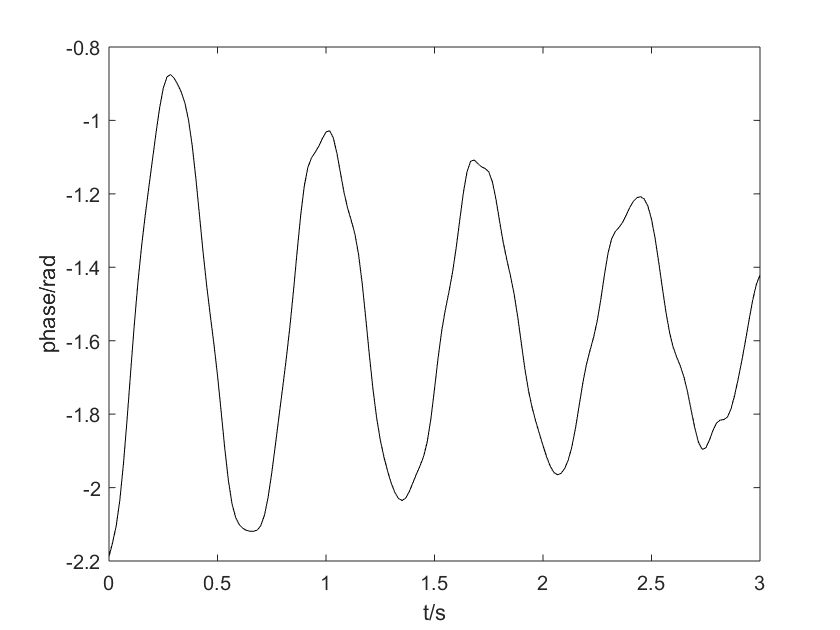}
    \caption{synthetic current phase}
\end{subfigure}
\begin{subfigure}[t]{.45\linewidth}
    \includegraphics[scale=0.2]{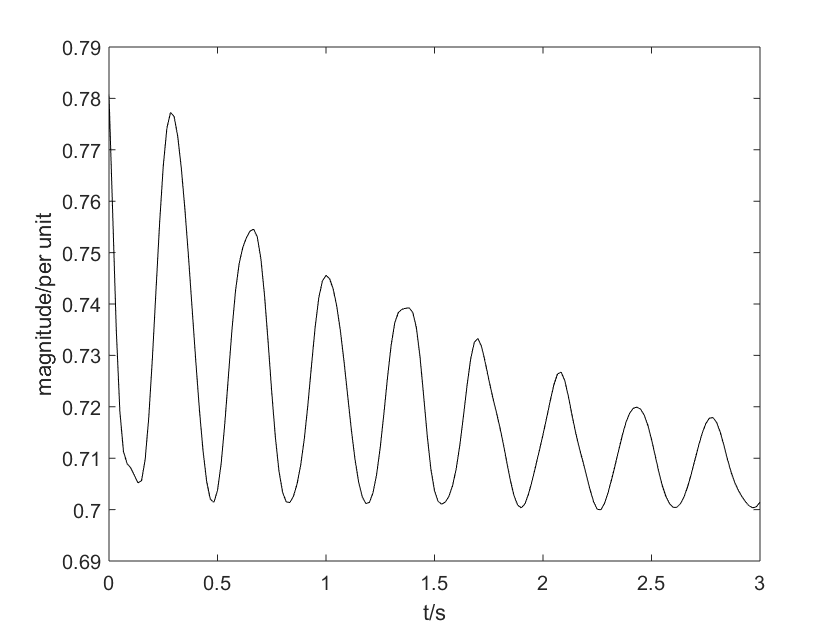}
    \caption{synthetic current magnitude}
\end{subfigure}
\caption{Comparison of real and synthetic samples in the SMIB system}\label{fig}
\end{figure}

\begin{figure}[t]
\centering
\begin{subfigure}[t]{0.45\linewidth}
    \includegraphics[scale=0.2]{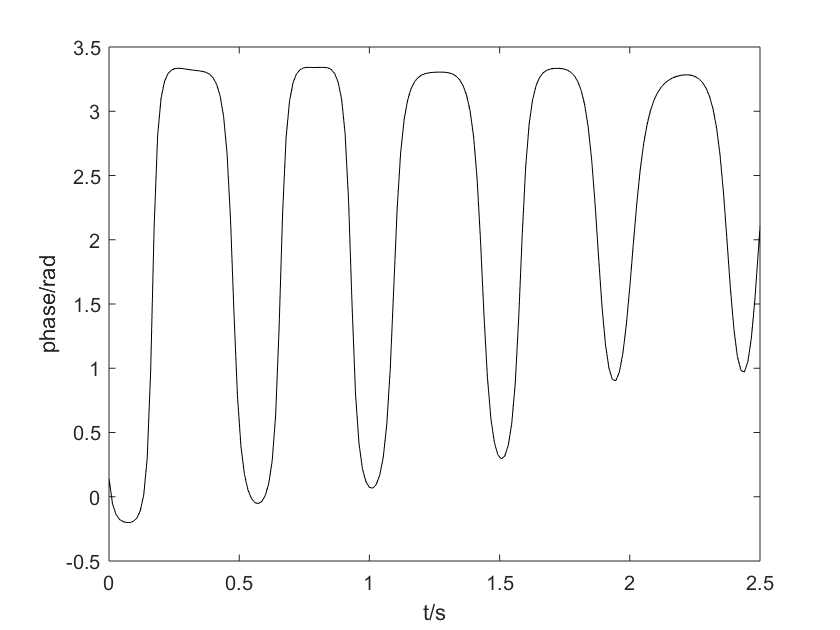}
    \caption{real current phase}
\end{subfigure}
\begin{subfigure}[t]{.45\linewidth}
    \includegraphics[scale=0.2]{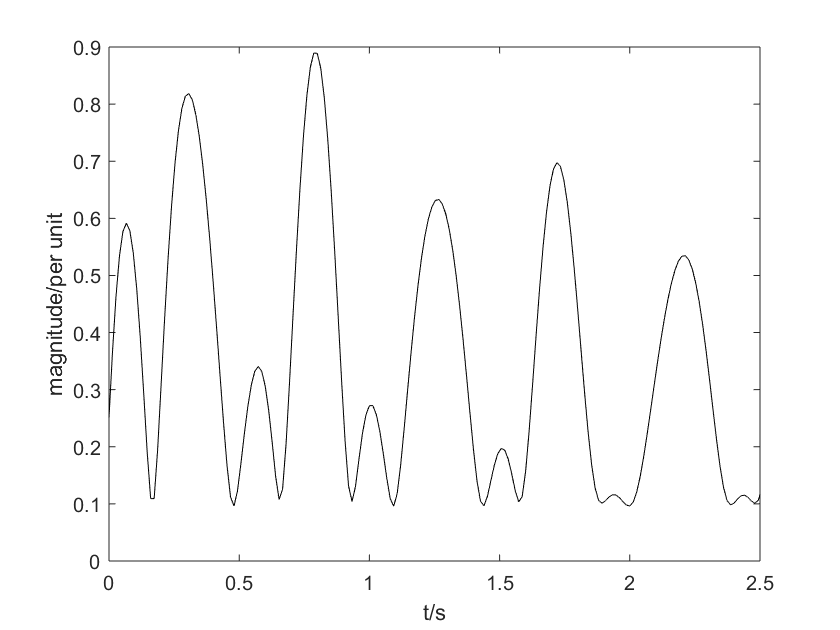}
    \caption{real current magnitude}
\end{subfigure}\\
\begin{subfigure}[t]{0.45\linewidth}
    \includegraphics[scale=0.2]{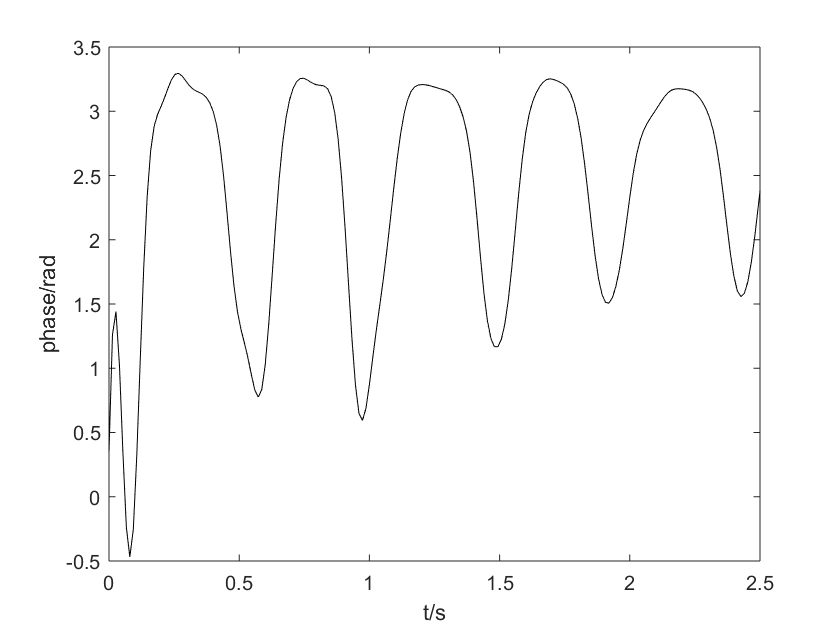}
    \caption{synthetic current phase}
\end{subfigure}
\begin{subfigure}[t]{.45\linewidth}
    \includegraphics[scale=0.2]{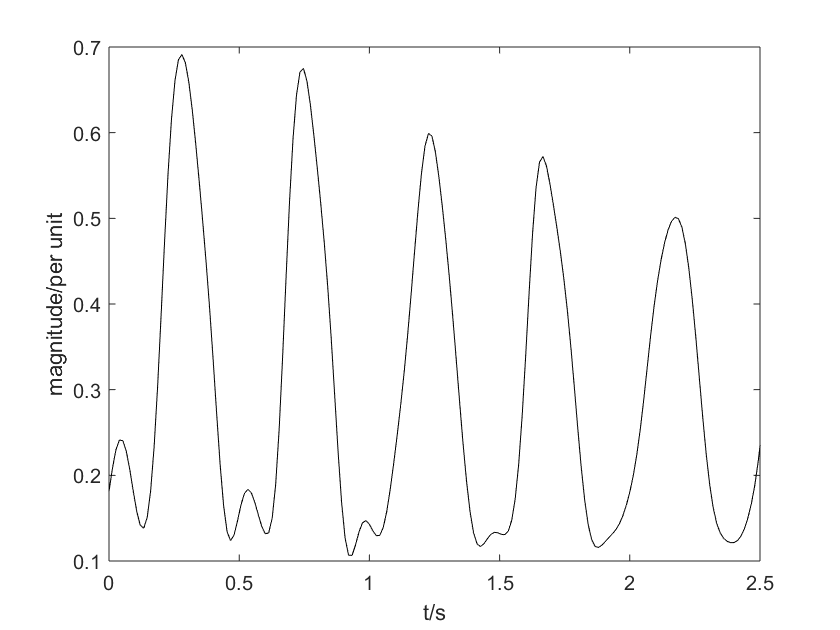}
    \caption{synthetic current magnitude}
\end{subfigure}
\caption{Comparison of real and synthetic samples in the IEEE 9-bus system}\label{fig}
\end{figure}

Figure 4 and 5 showed the comparison between real samples and selected synthetic samples by plotting the profiles out. Although the differences between real and synthetic samples are visible, a quantitative criterion is necessary to evaluate how realistic the synthetic samples are. From the perspective of power engineers, it is important to check whether the synthetic samples can be simulated from a physical model with the expected formula and how good the synthetic samples and simulated ones fit with each other. In the following part, a fitting method will be applied to identify the underlying SMIB system models regarding the synthetic samples. For IEEE 9-bus system, the complexity of its generator models brings more differential equations and undetermined parameters and hence it is difficult to infer an accurate model from the current and voltage measurements. Since such a problem is out of the scope of this paper, we only present the method how to identify the SMIB system model regarding the given current.

The basic is inspired by the literature\cite{b15}. The swing equation of the SMIB system can be written as:
\begin{equation}
\delta^{\prime\prime}+\alpha\delta^{\prime}+\beta{sin(\delta)}+\gamma=0
\end{equation}

As introduced in \cite{b15}, least square error method is applied to identify an equivalent SMIB model. Specifically, the parameters $\alpha$, $\beta$ and $\gamma$ are estimated by solving the fitting problem as shown in the following equation:
\begin{equation}
\begin{bmatrix}
\delta^{\prime}(t_1) & sin\delta(t_1) & 1 \\
\delta^{\prime}(t_2) & sin\delta(t_2) & 1 \\
\vdots & \vdots & \vdots\\
\delta^{\prime}(t_n) & sin\delta(t_n) & 1 \\
\end{bmatrix}
\begin{bmatrix}
\alpha \\ 
\beta \\ 
\gamma 
\end{bmatrix}
=
\begin{bmatrix}
-\delta^{\prime\prime}(t_1) \\ 
-\delta^{\prime\prime}(t_2) \\ 
\vdots \\ 
-\delta^{\prime\prime}(t_n) 
\end{bmatrix}
\end{equation}

Suppose all the fictitious SMIB models regarding the synthetic samples share the same network parameters. Then the profiles of the generator rotor angle can be obtained according to the synthetic samples. After applying the proposed method, the identified physical model can simulate the profile of the generator rotor angle with the same initial state of the synthetic profile. The realistic degree of synthetic samples can be evaluated by computing the mean relative error between model-derived profile and synthetic one as defined in the following equation, where $\delta$ is the synthetic profile of the generator inner phase and $\delta_{est}$ is the model-derived profile. The more realistic sample, the smaller error.

\begin{equation}
Error=mean\left( \frac{\vert\delta_{est}-\delta\vert}{max(\delta)-min(\delta)}\right) 
\end{equation}

\begin{figure}[t]
\centering
\includegraphics[scale=0.4]{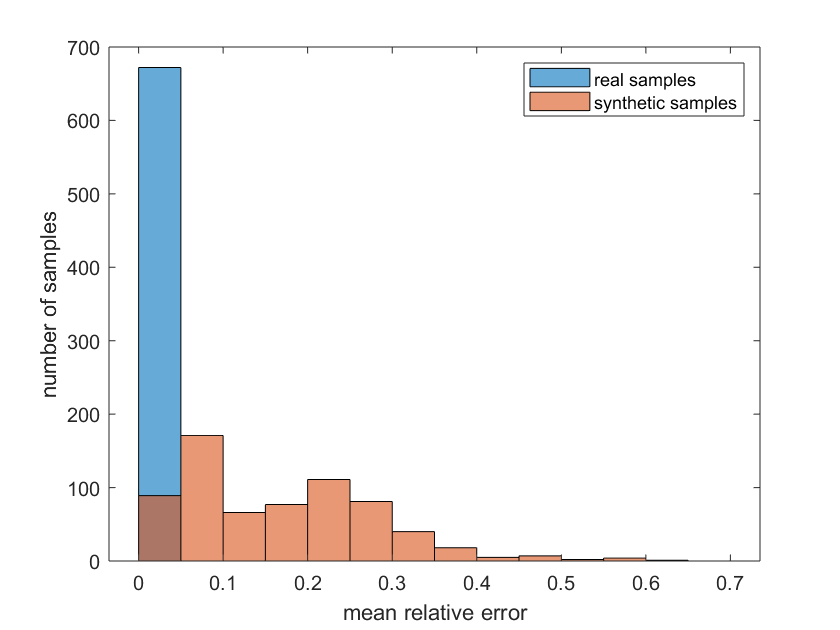}
\caption{The distribtuion of mean relative error of real and synthetic samples.}\label{fig}
\end{figure} 

Figure 6 shows the distribution of mean relative error of real and synthetic samples. The error of real samples concentrate near 0, while the error of synthetic samples spread over a relatively wide range. The fact that the error of real samples are close to 0 verifies the reasonability of this evaluation method. It is clear that only part of synthetic samples are realistic enough according to this evaluation method. Therefore an threshold is needed to determine whether the given sample can be regarded as a realistic sample. Figure 7 illustrates the synthetic and model-derived rotor angle profiles with different errors, where red curves represent the synthetic profiles and blue curves represent the model-derived profiles. According to the visible comparison, we classify a sample as unrealistic one if the mean relative error is over 0.09. Under such threshold, 36.16$\%$ of the synthetic samples are realistic. The result can be interpreted from two perspectives: (1) GAN has the capability of generating PMU profiles that follow some physical model; (2) The potential of GAN still has much rooms to explore.

\begin{figure}[t]
\centering
\begin{subfigure}[t]{0.45\linewidth}
    \includegraphics[scale=0.2]{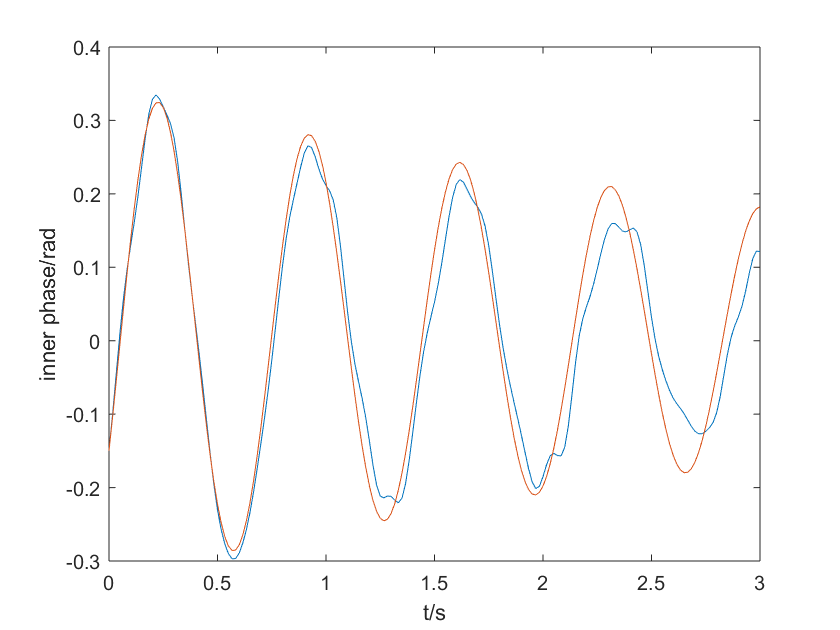}
    \caption{error=0.05}
\end{subfigure}
\begin{subfigure}[t]{.45\linewidth}
    \includegraphics[scale=0.2]{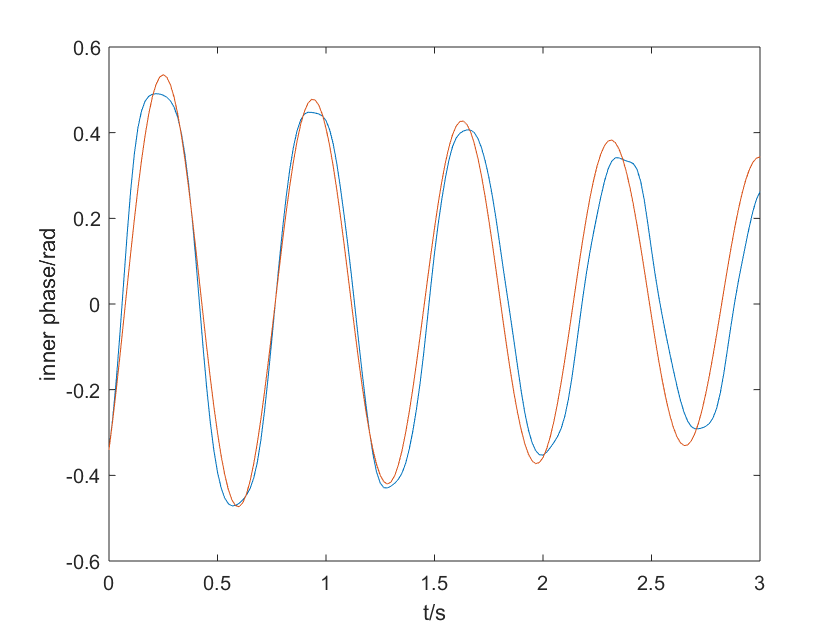}
    \caption{error=0.07}
\end{subfigure}\\
\begin{subfigure}[t]{0.45\linewidth}
    \includegraphics[scale=0.2]{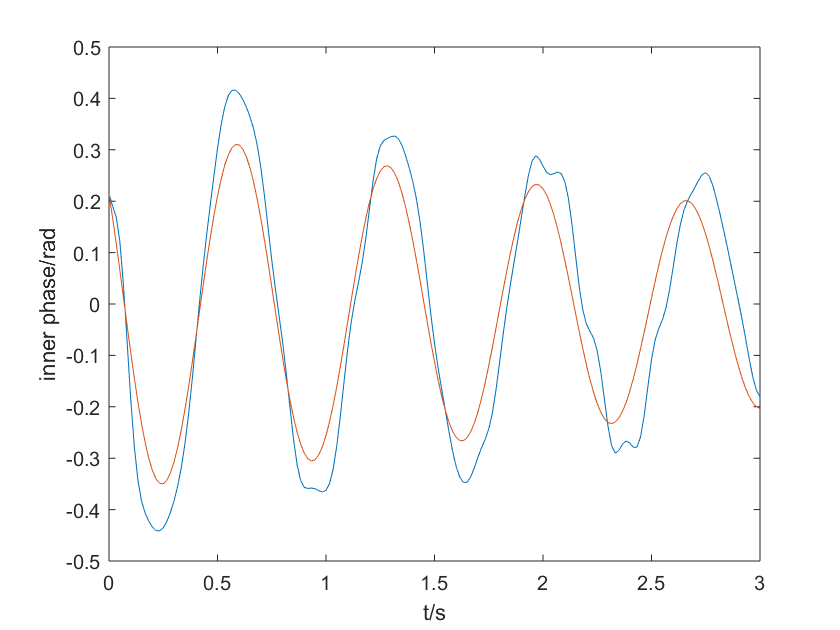}
    \caption{error=0.09}
\end{subfigure}
\begin{subfigure}[t]{.45\linewidth}
    \includegraphics[scale=0.2]{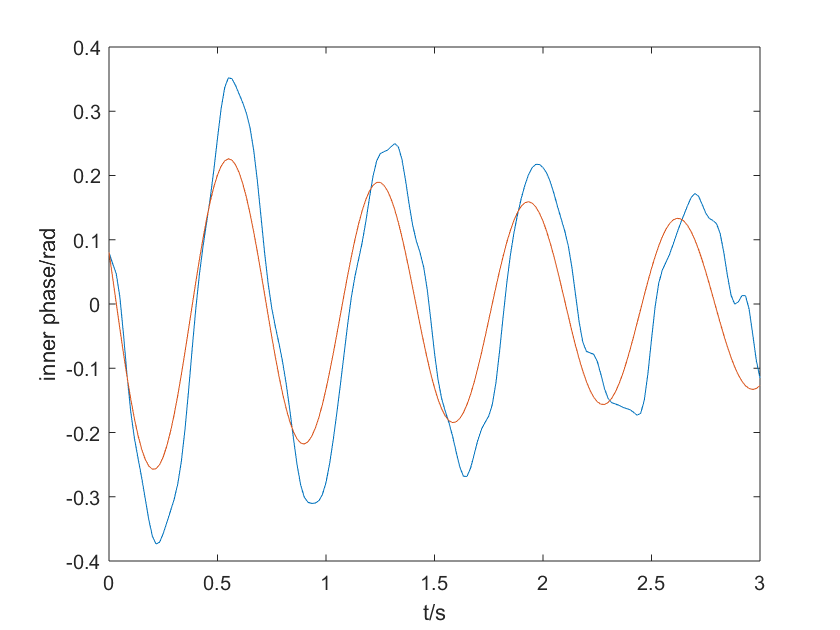}
    \caption{error=0.11}
\end{subfigure}
\caption{Illustration of synthetic inner phases with different errors}\label{fig}
\end{figure}

\section{Conclusion and Future Work}
In this paper, aiming to synthesize the realistic PMU data without using any physical model, we propose to use GAN to achieve this. We introduce the basic idea of GAN, an innovative generative model, and further clarify the objective function of Wasserstein GAN with the detailed training procedure. Then, applying Wasserstein GAN to two different power systems shows its potential to synthesize realistic PMU time series. However, through the quantitative evaluation by a proposed evaluation method, it is revealed that only part of synthetic samples are realistic enough, which means that there is still a large room for improvement.

During the experiment, it is shown that several aspects deserve further investigation: (1) The synthetic profiles has limited diverse scenarios, which means there might exist mode collapse during the training. Therefore it would be necessary to apply techniques to prevent this phenomenon if that is the case; (2) The input of Wasserstein GAN is random noise and hence what $G$ model mimics is not the dynamic function as discussed in the Section III. From the perspective of power engineering, it is much more desirable to have physical engineering interpretation. Considering this, recurrent neural networks might be an alternative for sequential generation, since its input naturally contain initial system states; (3) Once we could train GAN model to synthesize realistic PMU profiles for complex power systems, we need to figure out a general way to quantitatively measure the quality of the synthetic data.

\end{document}